\documentclass[letterpaper]{JHEP3}
\usepackage{cite}
\usepackage{graphicx}

\newcommand{\be}{\begin{equation}}
\newcommand{\ee}{\end{equation}}
\newcommand{\bea}{\begin{eqnarray}}
\newcommand{\eea}{\end{eqnarray}}
\renewcommand{\b}[1]{\bar{#1}}
\renewcommand{\t}[1]{\tilde{#1}}
\newcommand{\Del}{\nabla}
\newcommand{\del}{\partial}

\newcommand{\comment}[1]{}
\newcommand{\lb}[1]{\label{#1}}

\newcommand{\alp}{\alpha'}
\newcommand{\eq}[2]{\begin{equation}\label{#1}{#2}\end{equation}}
\newcommand{\normal}[1]{:\! #1 \!:}
\newcommand{\ket}[1]{| #1 \rangle}

\title{Backreaction in Closed String Tachyon Condensation}
\author{Andrew R. Frey\\ Department of Physics\\ McGill University\\
Montr\'eal, QC H3A 2T8, Canada\\
\email{frey@hep.physics.mcgill.ca}}

\abstract{We consider backreaction due to production of massless strings
in the background of a condensing closed string tachyon.  Working in
the region of weak tachyon, we find the modified equations of motion for
massless strings with conformal perturbation theory.  We solve for the
positive and negative frequency modes and estimate the backreaction on
the background dilaton.  In large (supercritical) dimensions, we find
that the backreaction can be significant in a large region of spacetime.
We work with the bosonic string, but we expect these results to carry over
into the heterotic case.
}

\preprint{}

\begin{document}

\section{Introduction}\lb{s:intro}

Tachyon condensation in various string constructions 
has become increasingly important as a set of clean 
examples of time-dependent systems in string theory and as a dynamical 
connection between varying states of string theory.  The most famous 
example is condensation of the open string tachyon on either an unstable
D-brane or a D-brane/anti-D-brane pair; Sen's conjecture, which has by now
considerable supporting evidence, states that tachyon condensation annihilates
the branes and leaves a purely closed string state behind (see
\cite{hep-th/0102085,hep-th/0109182,hep-th/0311017,hep-th/0410103} for
reviews).  In fact, recent work has found an exact dynamical 
solution connecting the unstable brane system to the closed string vacuum 
in string field theory \cite{arxiv:0803.1184}.

Closed string tachyons have also generated a great deal of interest, 
beginning with tachyons localized at special points in space
\cite{hep-th/0108075,hep-th/0111051,hep-th/0111154,hep-th/0111004,hep-th/0306146,hep-th/0312213,hep-th/0405064,hep-th/0502021,hep-th/0503073,hep-th/0403051,hep-th/0506130,hep-th/0506166,hep-th/0601032}.
For example, tachyons can develop at orbifold fixed points with conical
singularities; tachyon condensation appears to reduce the rank of the orbifold
group and eventually resolve the singularity.

Of course, bosonic string theory (and various nonsupersymmetric heterotic
theories) has a nonlocalized ``bulk'' closed string tachyon.  
Bulk closed string tachyon condensation seems to realize the conjecture of
\cite{Horowitz:2000gn,DeAlwis:2002kp,hep-th/0602230} that the closed string
tachyon ``vacuum'' is a bubble of nothing along the lines of 
\cite{Witten:1981gj}.  Bulk tachyon condensation has been studied in 
\cite{hep-th/0405041,hep-th/0408124,hep-th/0409071,hep-th/0506077,hep-th/0510126,hep-th/0611317,hep-th/0612051,hep-th/0612116,hep-th/0612031,arxiv:0709.2162,arxiv:0709.2166,arxiv:0709.3296,arxiv:0804.0697};
roughly speaking,
the tachyon generates a mass for some of the worldsheet fields, effectively
turning off propagation and oscillation of strings in the corresponding
dimensions of spacetime.  Therefore, as the tachyon condenses, spacetime loses
dimensionality; central charge is unchanged during tachyon condensation
because the dilaton background changes at the same time \cite{hep-th/0409071}.
Perhaps the natural tachyon profile to consider is exponential growth in a
timelike direction; these tachyon profiles have $\alp$ corrections that are
suppressed in supercritical string theories with large dimensionality $D$
\cite{hep-th/0409071}.  However, 
the cleanest results have been achieved in backgrounds in which the tachyon
grows exponentially along a lightlike direction; this tachyon profile is
an exact solution of perturbative string theory at all orders in $\alp$
\cite{hep-th/0611317,hep-th/0612051,hep-th/0612116}.  (Of course, there are
still corrections at nonzero string coupling.)

Compared to these impressive results in worldsheet physics, our understanding
of closed string tachyon condensation in spacetime
is somewhat behind.  Some progress has
been made in developing the effective field theory of massless closed strings
and tachyons, including some solutions of that theory
\cite{hep-th/0506076,hep-th/0702147,arXiv:0705.3265}, though a fully consistent
action has only been described recently \cite{arxiv:0804.2262}.
Nonetheless, it has been proposed that tachyon condensation could provide a 
stringy resolution of cosmological singularities 
\cite{hep-th/0506130,hep-th/0606127}.  This proposal was taken a step farther
in \cite{arXiv:0706.1104}, which assumed the Big Bang could be replaced by 
emergence from a tachyon condensate and asked what quantum mechanical
perturbations are generated by the time dependence of the tachyon; these
fluctuations could potentially serve as the initial values for 
inflationary perturbations.  That 
paper used a somewhat heuristic worldsheet Hamiltonian approach to calculate
particle production and found that the late-time fluctuations are a thermal
state at the energy scale set by the tachyon gradient.\footnote{Cosmological
questions relevant to open-string tachyon condensation have recently been
addressed in \cite{Aref'eva:2008gj}.}

In this paper, we reconsider particle production in a background of 
tachyon condensation, working in the usual picture that the tachyon grows 
toward the future.  Our goal is to determine if quantum fluctuation of 
massless string states, in particular the dilaton, backreact significantly on
known classical tachyon backgrounds.  We proceed by using conformal 
perturbation theory to determine how massless strings propagate in a weak
tachyon background. We then solve the modified equation of motion and 
calculate the quantum source for the dilaton (analogous to finding the quantum
mechanical stress-energy tensor in a cosmological background).  We find
that backreaction can be large even when most $g_s$ corrections are small.
Although we focus on backreaction in this paper, the reader should note that
the amplitude calculation we have done (and similar calculations) are 
useful for confirming the effective action of \cite{arxiv:0804.2262}.

The plan of this paper is as follows.  In section \ref{s:review}, we review
the tachyonic backgrounds of bosonic string theory that we consider.  We
also show directly that the lightlike tachyon background is a conformal field
theory on the worldsheet.\footnote{We thank S.~Hellerman for his patient
explanation of this calculation.}  Then, in section \ref{s:propagation},
we calculate the scattering amplitude of a tachyon and two massless string
states.  (The appendix contains a review of the BRST quantization of the
string in the linear dilaton background, which is useful for the calculation
of this amplitude.)
In conformal perturbation theory, this amplitude gives the modified 
propagator of the massless states in a weak tachyon background, which 
directly tells us the modified equation of motion for perturbations of the
massless fields.  Finally, in section \ref{s:backreaction}, we calculate
the quantum mechanical source for the dilaton, which we show to be large in
many circumstances.

\section{Review of Bulk Tachyon Condensation}\lb{s:review}

In this section, we review known results about closed string tachyon
condensation from the worldsheet perspective.  
Rather than consider tachyons localized at some point in the 
spacetime (such as a shrinking circle or a nonsupersymmetric orbifold fixed
point), we will restrict our attention to the bulk closed string tachyons
of the bosonic theory. We will mostly discuss tachyons 
with a lightlike gradient as
discussed in \cite{hep-th/0611317,hep-th/0612051,hep-th/0612116}, since they 
are uncorrected in $\alp$, but we will also show how our discussion carries
over for tachyons with a timelike gradient.

We begin by reminding the reader of allowed tachyon backgrounds in 
supercritical string theory, largely following 
\cite{hep-th/0409071,hep-th/0611317}.  Then we show explicitly that the 
conformal anomaly vanishes for the lightlike tachyon gradient.  As far as
we know, this proof has been known for some time but has not appeared in 
the literature \cite{hellermanprivate}.

A brief note on conventions: we take lightcone coordinates so that the 
Minkowski metric is
\eq{minkowski}{ds^2 = -(dX^0)^2+(dX^1)^2+d\vec X^2 =
-2dX^+ dX^- +d\vec X^2\ .}

\subsection{Tachyon Vertex Operator}

In order to control worldsheet corrections to tachyon condensation, we work
in a linear dilaton background, typically in large supercritical dimension
$D$.  (In fact, we will typically consider the tachyon to be a small 
perturbation to the linear dilaton.)  In the string frame (\textit{i.e.}, 
the spacetime fields that couple to the string variables), the metric is
Minkowski and the dilaton takes the form $\Phi=V_\mu X^\mu$ with 
$V^2=(26-D)/6\alp$.  The linear dilaton background is a well-known exact
solution of tree-level string theory \cite{Polchinski:1998rq}.  We choose
the dilaton to decrease into the future, $V_0 = -\sqrt{(D-26)/6\alp}$.
We can also consider a tachyon background in the critical dimension $D=26$,
in which case we take only $V_-$ to be nonzero; typically, though, we 
will work in supercritical dimension.

As has been observed, for example in \cite{hep-th/0409071},
the condition for a tachyon vertex operator to have
the correct weight in a general linear dilaton background is
\eq{onshell}{
\del_\mu\del^\mu T(X) -2V^\mu \del_\mu T(X)+\frac{4}{\alp}T(X)=0\ .}
Due to the tachyonic mass term, the solution will have some time dependence,
which we take to be exponential in either $X^0$ or $X^+$.

It is easy to check that 
\eq{nospace}{T(X)=\frac{\mu_0^2}{2} e^{\beta X^+}}
solves (\ref{onshell}) for $\beta=2/V^+\alp$.  In most string theories,
this solution represents a bubble of nothing.\footnote{As was explained in 
\cite{hep-th/0612116}, this tachyon background beginning in the type 0
string is a transition to the bosonic string theory.}
Of more interest to us
is a tachyon with some spatial dependence,
\eq{tach1}{T(X) = \frac{\mu^2}{2\alp}e^{\beta X^+}\!\!
\normal{(X^a)^2}+\delta T(X)\ .}
Here, $(X^a)^2$ is a sum over $N$ of the transverse spatial
directions (rather than all the spatial dimensions).  
(The reader should note that these are \textit{transverse} dimensions,
so we necessarily have $N\leq D-2$.)
We include $\delta T$ in order
to solve the equation of motion; then we find
\eq{tach2}{\delta T(X) = \frac{N\mu^2}{4}\left(\beta X^+\right)
e^{\beta X^+}}
for the specific solution (of course, (\ref{nospace}) can be added 
independently), as explained in \cite{hep-th/0612051}.

The tachyon with quadratic spatial dependence (\ref{tach1}) can be written
as the long-wavelength limit of a plane wave, as in \cite{hep-th/0612051}.
In particular, 
\eq{planetach}{T(X) = \mu^2 e^{\beta_k X^+}\!\!
\normal{e^{i\vec k\cdot \vec X}}\ ,\ \ V^+\beta_k =\frac{2}{\alp}-
\frac{1}{2}\vec k^2}
is a valid tachyon background.  Without loss of generality, we can take
$\vec k$ to point along a single axis (say the $Y$ direction), in which
case (\ref{tach1},\ref{tach2}) (with $N=1$) 
follow by taking the $k\to 0$ limit of 
\eq{costach}{T(X) = \mu_0^2 \left(e^{\beta X^+} - e^{\beta_k X^+}\!\!
\normal{\cos(kY)}\right)}
with $\mu^2= \alp\mu_0^2 k^2$ fixed.  Incidentally, (\ref{planetach}) shows 
immediately that, even in the critical dimension, the lightlike tachyon 
only solves the equations of motion in a linear dilaton background.

The story is very similar with a timelike tachyon gradient 
\cite{hep-th/0409071}.  The tachyon profile at fixed momentum is
\eq{planetachT}{T(X) = \mu^2 e^{\beta_k X^0}\!\!
\normal{e^{i\vec k\cdot \vec X}}\ ,\ \ \beta_k^2-2V_0\beta_k =\frac{4}{\alp}
-\vec k^2\ .}
In the large $D$ limit, we can drop the first term in the quadratic equation
for $\beta_k$, finding exactly the same dependence as in (\ref{planetach}).
Again, as long as we work in the large dimension limit, we can take
the same $\vec k\to 0$ limit to find a background
\eq{quadtachT}{T(X) = \frac{\mu^2}{2\alp}e^{\beta X^0}\!\!
\normal{(X^a)^2}+\frac{N\mu^2}{4}\left(\beta X^0\right)
e^{\beta X^0}\ ,}
where $\beta V_0 = -2/\alp$.   These tachyon backgrounds do have $\alp$ 
corrections, but those corrections are suppressed at large $D$.

Finally, the reader should note that the tachyon ``amplitude'' $\mu^2$ can
be either positive or negative sign, depending on which way the tachyon rolls
off its local maximum.  We have chosen to write the amplitude as a squared
mass to make dimensional factors work out more easily.

\subsection{Conformal Invariance}\lb{s:conformal}

We now remind the reader that the interacting 
worldsheet theory in the presence of the tachyon is actually a conformal
field theory for the lightlike tachyon gradient.  
It seems that this proof
has been known for some time but has not yet made its way into the 
literature \cite{hellermanprivate}.  For simplicity, we work with the
$\vec k\to 0$ limit of (\ref{tach1},\ref{tach2}).

To get a string background with a nontrivial tachyon, we can (following
the usual procedure) just exponentiate the tachyon vertex operator in
the linear dilaton CFT.  The Polyakov action then becomes
\bea
S&=& -\frac{1}{4\pi\alp} \int d^2\sigma\sqrt{-\gamma} \left[
\del_\alpha X^\mu\del^\alpha X_\mu +\frac{\mu^2}{2}e^{\beta X^+}\left(
(X^a)^2+\frac{N\beta\alp}{2}X^+\right)\right]\nonumber\\
&&+\frac{\{i\}}{4\pi}\left(\int d^2\sigma \sqrt{-\gamma}R V_\mu X^\mu
+2\int_{\del M}d\sigma K V_\mu X^\mu\right)\ .\lb{polya}\eea
We have included the contribution from the linear dilaton just to remind
the reader that it is there; the 
worldsheet boundary includes ``boundaries at infinity.''  At this stage,
it doesn't matter whether we are using a Euclidean or Lorentzian worldsheet;
the factor of $i$ in curly braces is appropriate for the Lorentzian case.

\FIGURE{
\includegraphics[scale=1]{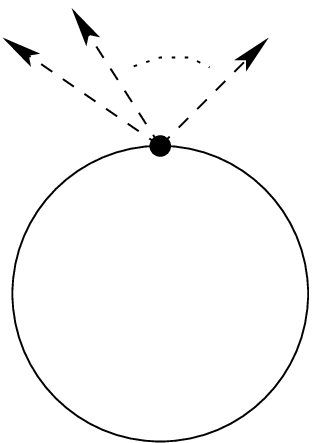}
\caption{\lb{f:loop}The only diagrams contributing divergences are those
with a single vertex with an arbitrary number of $X^+$ lines leaving and
one $X$ propagator in a loop.}}

Let us make more explicit the 
correct renormalized (UV finite) form of the action.  Beyond the normal
divergences of the free theory, the tachyonic theory only gets UV divergences
from loops with a single propagator (from \cite{hep-th/0612051}, we know
that there are no diagrams beyond one loop), as illustrated in figure 
\ref{f:loop}.  With a point-splitting regulator, the divergence comes from
the coincidence of the two ends of the propagator.  In this limit, then,
the divergence is the same as in $\mathbb{R}^{1,1}$, namely
\eq{diverge}{
-\frac{N \alp\mu^2}{4} e^{\beta X^+} \ln\left(\Delta s^2\right)\ ,}
where $\Delta s^2$ is the proper distance between the two ends of the
propagator (we have chosen the coordinate invariant form for obvious reasons).

To write the renormalized action, then, we only need to replace $(X^a)^2$ with
the conformally normal ordered version $\normal{(X^a)^2}$, 
just as we would guess from
the operator treatment.  (Note that the free action should technically replace
$(\del X)^2$ similarly with $\normal{(\del X)^2}$).  This renormalized
action is the proper starting point for the Hamiltonian treatment described in 
\cite{arXiv:0706.1104}.

By taking the variation of the action with respect to the metric, we find
the stress tensor
\bea
T_{\alpha\beta} & =& -\frac{1}{\alp} \left(\normal{\del_\alpha X^\mu
\del_\beta X_\mu}
-\frac{1}{2}\gamma_{\alpha\beta}\normal{\del_\gamma X^\mu\del^\gamma X_\mu}
\right)
+V_\mu\Del_\alpha\Del_\beta X^\mu-\gamma_{\alpha\beta}
V_\mu\Del^2X^\mu\nonumber\\
&&+\frac{\mu^2}{4\alp}\gamma_{\alpha\beta}\, 
e^{\beta X^+}\left(\normal{(X^a)^2}+\frac{N\beta\alp}{2} X^+\right)
+\frac{N \mu^2}{4} e^{\beta X^+}
\frac{\Delta\sigma_\alpha\Delta\sigma_\beta}{\Delta s^2}\ .
\lb{stress}
\eea
The first line of (\ref{stress}) is the usual contribution from the kinetic
term and the linear dilaton, while the second line gives the classical 
and quantum contributions from the tachyon (potential) term.  Note that the
last, quantum mechanical, term arises from the metric dependence in 
$\normal{X^2}$.  When properly averaged over possible directions, 
$\Delta\sigma_\alpha\Delta\sigma_\beta
\to (1/2)\gamma_{\alpha\beta}\Delta s^2$.  The reader 
should also be aware that a similar quantum mechanical term should arise 
from the normal ordering of the kinetic terms, but it cancels out.

Conformal invariance is simple enough to demonstrate.  The trace of the
stress tensor is just
\eq{trace}{T^\alpha_\alpha = -V_- \Del^2 X^- +\frac{\mu^2}{2\alp}e^{\beta X^+}
\left(\normal{(X^a)^2}+\frac{N\beta\alp}{2}X^+ + \frac{N}{2}\right)\ .}
We can simplify this expression using the equation of motion for $X^-$,
which is 
\eq{eomminus}{
2\Del^2 X^- = -\frac{\beta\mu^2}{2} e^{\beta X^+} \left(\normal{(X^a)^2}
+\frac{N\beta \alp}{2}X^+ +\frac{N\alp}{2}\right)\ .}
Along with $V_- = -2/\beta\alp$, this yields $T^\alpha_\alpha=0$.

In terms of worldsheet lightcone coordinates on a flat worldsheet, we find
the usual stress tensor 
\bea
T_{++} &=& -\frac{1}{\alp}\normal{\del_+ X^\mu\del_+ X_\mu} + V_-\del_+^2 X^-
\nonumber\\
T_{--} &=& -\frac{1}{\alp}\normal{\del_- X^\mu\del_- X_\mu} + V_-\del_-^2 X^-
\ .\lb{stressflat}\eea
As usual, the central charge from the total dimensionality, linear dilaton,
and conformal ghosts cancel.  The tachyon does not contribute to the
central charge at all.

\section{Propagation in the Weak Tachyon Region}\label{s:propagation}

In this section, we compute the string perturbation theory amplitude for 
a single tachyon and two massless strings in a linear dilaton background.
This amplitude provides the first order correction to the massless string
action in the tachyon background.  We can then see how the massless mode
propagation differs in the presence of a weak tachyon.  In this section,
we work on a Euclidean worldsheet with complex coordinates $z$, and we 
work at tree level in string perturbation theory.

\subsection{Tachyon Modification to Action}\label{s:amplitude}

The propagation of massless string modes will be modified by scattering off
the tachyon background; the one-tachyon/two-massless-string amplitude gives 
the shift in the two-point massless correlation function and therefore the
action for the massless strings.  Since we want to think of massless string
scattering from the tachyon background, at lowest order we should calculate
the string diagram with three vertex operators.
\comment{given in figure \ref{f:source}.}  
(Higher order calculations
in the tachyon amplitude correspond to adding more tachyon vertex operators.)

The two massless string vertex operators are given by 
\eq{masslessVO}{e^{1,2}_{\mu\nu}\normal{\del X^\mu\b\del X^\nu 
e^{ik_{1,2}\cdot X}}(z_{1,2},\b z_{1,2})\ ,}
in which the polarization tensors and momenta obey the gauge conditions given
in (\ref{masslesscohomology}).  In particular, the mass-shell and gauge
conditions are
\eq{msgauge}{k_{1,2}\cdot(k_{1,2}+2iV)=0\ ,\ \ 
e^{1,2}_{\mu\nu}(k_{1,2}+2iV)^\nu = e^{1,2}_{\nu\mu}(k_{1,2}+2iV)^\nu=0\ .}
For ease of calculation, we will use the plane-wave tachyon vertex operator
(\ref{planetach}), and we will combine the $X^+$ and $X^i$ exponentials into
a single exponential with relativistic momentum $k^3_\mu$.  
We will not concern ourselves with the overall scaling of the vertex operators,
instead normalizing the final shift of the action to get the correct
dimensionality.

Since
there are only three vertex operators, we work with the fixed-position form
of the amplitude (up to normalization)
\eq{ampdef}{\mathcal{A}=
e^1_{\mu\nu}e^2_{\lambda\rho} \left\langle \normal{\del X^\mu\b\del X^\nu 
e^{ik_{1}\cdot X}}(z_{1},\b z_{1}) \normal{\del X^\lambda\b\del X^\rho
e^{ik_{2}\cdot X}}(z_{2},\b z_{2}) \normal{e^{ik_3\cdot X}}(z_3,\b z_3)
\right\rangle\times(\textnormal{ghosts})\ .}
The vertex operator positions $z_i\in\mathbb{C}$ are arbitrary fixed positions
on the worldsheet.
The path integral over the $X^\mu$ zero modes gives a momentum preserving
delta-function; however, in the linear dilaton background, the sum of 
momenta also includes the dilaton gradient $V^\mu$ (see, for example, 
exercises in \cite{Polchinski:1998rq}).  Therefore, we have
\eq{delta}{\mathcal{A}\propto (2\pi)^D\delta^D\left(\sum_i k_i +2iV\right)\ .}
The ghost contribution is unchanged from that in the Minkowski background
of critical string theory.
The expectation value can be simplified through the use of (\ref{msgauge})
to read 
\bea
\mathcal{A}&=& (2\pi)^D\delta^D\left(\sum_i k_i +2iV\right)
e^1_{\mu\nu}e^2_{\lambda\rho}\left|z_{12}\right|^{\alp k_1\cdot k_2}
\left|z_{13}\right|^{\alp k_1\cdot k_3}\left|z_{23}\right|^{\alp k_2\cdot k_3}
\nonumber\\
&&\times\left(\eta^{\mu\lambda}-\frac{\alp}{8}k_{23}^\mu k_{13}^\lambda\right)
\left(\eta^{\nu\rho}-\frac{\alp}{8}k_{23}^\nu k_{13}^\rho\right)\ ,
\lb{amp1}\eea
where $z_{ij}=z_i-z_j$ and similarly for $k_{ij}$.

We get the physically meaningful amplitude (again, up to dimensionful 
normalizations) by using a coordinate transformation to choose values of 
the vertex operators.  A particularly convenient choice is $z_1=0$,
$z_2=1$, $z_3\to\infty$; this limit is well defined due to momentum 
conservation and the appropriate mass-shell conditions.  We end up with
\eq{amp2}{\mathcal{A}=(2\pi)^D\delta^D\left(\sum_i k_i +2iV\right)
e^1_{\mu\nu}e^2_{\lambda\rho}\left(\eta^{\mu\lambda}-
\frac{\alp}{8}k_{23}^\mu k_{13}^\lambda\right)
\left(\eta^{\nu\rho}-\frac{\alp}{8}k_{23}^\nu k_{13}^\rho\right)\ .}

In the effective theory of the massless strings, this amplitude just
corresponds to a shift in the quadratic part of the action.  We will 
specialize to the graviton and dilaton.  In that case, the polarization
tensor becomes
\eq{polarization}{e_{\mu\nu}^{1,2} = h_{\mu\nu}(k_{1,2}) +\gamma\phi(k_{1,2})
\eta_{\mu\nu}\ ,}
where $\gamma$ is a numerical coefficient.  We can now Fourier transform
back to spacetime to get the change in the quadratic 
action in the presence of a tachyon background.  We keep all terms up to
second order in derivatives and find
\bea
\Delta S &=& \mu^2 \int d^Dx\, e^{-2V\cdot X}e^{ik\cdot X}\left[
4\left(h_{\mu\nu}+\gamma\phi\eta_{\mu\nu}\right)\left(h^{\mu\nu}
+\gamma\phi\eta^{\mu\nu}\right)+\alp(\del+ik)^\nu h_{\mu\lambda}
(\del+ik)^\mu h_\nu{}^\lambda\right.\nonumber\\
&&\left.+2\gamma\alp(\del+ik)^\mu h_{\mu\nu}(\del+ik)^\nu \phi
+\gamma^2\alp(\del+ik)_\mu\phi (\del+ik)^\mu\phi\right.\nonumber\\
&&\left.-\frac{1}{4}\alp{}^2
k^\lambda(\del+ik)^\nu h_{\mu\lambda} k^\rho (\del+ik)^\mu h_{\nu\rho}
-\frac{1}{2}\gamma\alp{}^2 k\cdot (\del+ik)h_{\mu\nu}k^\mu (\del+ik)^\nu\phi
\right.\nonumber\\
&&\left.
-\frac{1}{4}\gamma^2\alp{}^2 k\cdot (\del+ik)\phi k\cdot (\del+ik)\phi
-\frac{1}{8}\alp{}^2 h_{\mu\nu}(\del+ik)^\mu (\del+ik)^\nu h_{\lambda\rho}
k^\lambda k^\rho \right.\nonumber\\
&&\left. -\frac{1}{8}\gamma \alp{}^2k^2 h_{\mu\nu}(\del+ik)^\mu (\del+ik)^\nu
\phi -\frac{1}{8}\gamma \alp{}^2 \phi (\del+ik)^2 h_{\mu\nu}k^\mu k^\nu
\right.\nonumber\\
&&\left. -\frac{1}{8}\gamma \alp{}^2 k^2 \phi (\del+ik)^2\phi
-\frac{i}{4}\alp{}^2 h_{\mu\nu}k^\mu (\del+ik)^\nu h_{\lambda\rho}k^\lambda
k^\rho-\frac{i}{4}\gamma\alp{}^2\phi k\cdot (\del+ik) h_{\mu\nu}k^\mu k^\nu
\right.\nonumber\\
&&\left. -\frac{i}{4}\gamma\alp{}^2k^2 k^\mu h_{\mu\nu}(\del+ik)^\nu\phi
-\frac{i}{4}\gamma^2 \alp{}^2k^2\phi k\cdot(\del+ik)\phi
\right]\ .\lb{action}
\eea
We have restored prefactors, choosing them to get a dimensionless action; 
$h_{\mu\nu}$ and $\phi$ are canonically 
normalized in $D$ dimensions, and $\mu$ is the mass scale of the tachyon
background.  We have also dropped the subscript ``3'' on the tachyon momentum
for notational convenience.  Notice that the action (\ref{action}) is
complex; this is not surprising because we are so far working with a 
complex plane wave tachyon background.  Once we convert to a real tachyon
background, the action will be real.

\subsection{Modified Equations of Motion}\label{s:eom}

Since we are interested in the propagation of particles through the tachyon
background, we now turn to the equations of motion, which has an effect 
even at the linear level.  
Since the key physics is the same for both the graviton and the dilaton,
we will focus henceforth on the (slightly simpler) dilaton equation of
motion.  In addition, we will set graviton fluctuations to 
zero from this point; while dilaton fluctuations necessarily source graviton
fluctuations, this mixing is not new to tachyonic backgrounds.  Part of the
mixing of the dilaton and graviton is due to working in string frame, and part
arises already in the pure linear dilaton background.  In order to focus on
the new physics, then, we just set the graviton fluctuations to zero.

The new contribution to the dilaton equation of motion from (\ref{action}) 
is then
\bea
\frac{\delta(\Delta S)}{\delta\phi} &=&\gamma^2
\mu^2 e^{ik\cdot X} e^{-2V\cdot X}
\left[8 D\phi -\alp  (\del-2V)\cdot(\del+ik) \phi\right.
\nonumber\\
&&\left.
+\frac{1}{2}\alp{}^2 k\cdot (\del-2V) k\cdot (\del+ik)\phi
-\frac{1}{8} \alp{}^2 k^2 (\del+ik)^2 \phi \right.\nonumber\\
&& \left. -\frac{1}{8}\alp{}^2 k^2 (\del-2V)^2 \phi
-\frac{i}{4} \alp{}^2  k^2 k\cdot (\del+ik)\phi
+\frac{i}{4} \alp{}^2 k^2 k\cdot (\del-2V) \phi\right]\ .\lb{dileom1}
\eea
We should now switch over to a real background for the tachyon.  
Since we are most interested in the time dependence of the
system, we will take the long wavelength limit of (\ref{tach1},\ref{tach2})
and (\ref{quadtachT}) for the lightlike and timelike tachyon gradients 
respectively.  For example, in the lightlike case, we set 
\eq{tachk}{k_+=-i\beta_k\ ,\ \ k_-=0\ ,\ \ k_i= k\delta_{ia}}
for a quadratic term in a specific direction $X^a$ and then sum over the
contributions for $N$ such directions.  In the following, we use subscripts $i$
to represent all (transverse) 
spatial dimensions, while subscripts $a$ represent only those
spatial dimensions on which the tachyon depends quadratically.

In the lightlike tachyon background (\ref{tach1},\ref{tach2}), the shift in the
dilaton equation of motion becomes
\bea
\frac{\delta (\Delta S)}{\delta\phi} &=& 2\gamma^2 T(X) e^{-2V\cdot X} \left\{
4D\phi +\alp\left[(\del_+ + \beta)(\del-2V)_-\phi +\del_-(\del-2V)_+\phi
-\del_i\del_i\phi\right]\vphantom{\frac{1}{4}}\right.\nonumber\\
&& \left. +\frac{1}{4}\alp{}^2\beta^2 \del_-(\del-2V)_-\phi\right\}
+\frac{1}{2}\gamma^2\alp\mu^2 e^{\beta X^+} e^{-2V\cdot X}\left\{-\del_a\del_a
\phi +\frac{1}{2}N\beta(\del-2V)_-\phi \right.\nonumber\\
&&\left. -\frac{1}{2}N\del_- \del_+\phi +\frac{1}{4}N\del_i\del_i\phi
+\frac{1}{4}(\del-2V)^2\phi \right\} \
-2\gamma^2\mu^2 e^{\beta X^+}e^{-2V\cdot X}X^a\left\{\del_a\phi
\vphantom{\frac{1}{4}}\right.\nonumber\\
&&\left. -\frac{1}{4}\alp\beta(\del-2V)_-\del_a\phi -\frac{1}{4}\alp\beta
\del_-\del_a\phi\right\}\ .\label{dileom2}
\eea
The timelike tachyon background (\ref{quadtachT}) yields a similar result,
which is not illustrative enough to repeat.  Now we note that the complete
dilaton equation of motion in the linear dilaton background is
\eq{PhiEOM}{\frac{\delta S}{\delta \Phi} =\sqrt{-g}e^{-2\Phi} \left[
-8V^2-2R+8\del_\mu\Phi\del^\mu\Phi-8 \Del^2\Phi\right]\ ,}
where $4V^2$ is the cosmological constant due to the supercritical dimension.
We write $\Phi=\Phi_0+V_\mu X^\mu +\kappa\phi$ and linearize to find the
equation of motion for fluctuations at zeroth order in the tachyon 
background.\footnote{$\kappa$, the $D$-dimensional Planck constant, is included
so that the fluctuation $\phi$ has canonical dimension (and normalization
up to constants of order unity).}  Combining, we find the equation of motion
to first order in the tachyon background, which is 
\bea
0 &= & \left(\del^2-2V\cdot \del\right)\phi -\frac{1}{4}\gamma^2 T(x)\left\{
4D\phi+\alp\beta(\del-2V)_-\phi +\frac{1}{4}\alp{}^2\beta^2\del_-
(\del-2V)_-\phi\right\}\nonumber\\
&&-\frac{1}{16}\gamma^2\mu^2\alp e^{\beta X^+}\left\{ \frac{1}{2}
N\beta (\del-2V)_-\phi -\del_a\del_a\phi +\frac{1}{2}NV\cdot\del\phi
-\frac{1}{2}V\cdot(\del-2V)\phi
\right\}\nonumber\\
&&+\frac{1}{4}\gamma^2\mu^2
e^{\beta X^+} X^a\del_a\left\{ \phi-\frac{1}{4}\alp\beta (\del-2V)_-\phi
-\frac{1}{4} \alp\beta\del_-\phi\right\}\lb{dileom3}
\eea
for the lightlike tachyon.  Note that, since we are working to first order in 
the tachyon, we were able to remove terms from (\ref{dileom2}) that are
proportional to the linearized (\ref{PhiEOM}).

For completeness, we also list the linearized dilaton equation of motion for
the timelike tachyon background.  It is
\bea
0&=& \left(\del^2 -2V\cdot \del\right)\phi -\frac{1}{4}\gamma^2 T(X) \left\{
4D\phi+\alp\beta(\del-2V)_0\phi +\frac{3}{8}\alp{}^2\beta^3(\del-2V)_0\phi
\right.\nonumber\\
&&\left. +\frac{3}{16}\alp{}^2\beta^4\phi +\frac{1}{2}\alp{}^2\beta^3 V_0\phi
+\frac{1}{4}\alp{}^2\beta^2 V_0^2\phi 
+\frac{1}{4}\alp{}^2\beta^2\del_0(\del-2V)_0\phi
\right\}\nonumber\\
&&-\frac{1}{16}\gamma^2\mu^2\alp e^{\beta X^0}\left\{ 
\frac{1}{2}N\beta(\del-2V)_0\phi -NV_0^2\phi -\frac{3}{2}N\beta^2\phi
-\del_a\del_a\phi -2N\beta V_0\phi
\right\} \nonumber\\
&&+\frac{1}{4}\gamma^2\mu^2 e^{\beta X^0}
X^a\del_a\left\{\phi-\frac{1}{8}\alp\beta^2\phi 
-\frac{1}{4}\alp\beta(\del-2V)_0\phi -\frac{1}{4}\alp\beta \del_0\phi\right\}
\ .\lb{dileom4}
\eea

Before moving on, we should pause to reflect on the various parts of 
(\ref{dileom3},\ref{dileom4}).  First of all, there are time dependent
mass terms, including contributions from the fact that the vertex operator
for $\phi$ includes polarizations in the $N$ spatial directions annihilated
by the tachyon.  Next, there are time dependent drag terms due to the tachyon 
gradient and an extra contribution due to momentum in the $X^a$ directions.
Also, there are terms with second order (lightcone) time derivatives, which
are reminiscent of the shift in the spacetime metric found in
\cite{hep-th/0409071,hep-th/0612051} through renormalization of the worldsheet
theory.  Finally, there are terms of the form $X^a P_a$.  Many of these
terms were anticipated by the Hamiltonian calculation carried out
in \cite{arXiv:0706.1104}.

\section{Backreaction}\label{s:backreaction}

We now have the pieces we need to study quantum backreaction effects in
tachyon condensation, working self-consistently in the region of spacetime
that has both small string coupling and small tachyon condensate.  We will
see that the backreaction due to quantum particle production can become
strong even in that region.

As a brief review, 
\cite{arXiv:0706.1104} studied particle creation in the
time-reversed background using an approximate equation of motion derived 
with a simplified Hamiltonian treatment of the string, concluding
that the tachyon ``decondensation'' resulted in a thermal bath at 
inverse temperature $\beta$, which is about the string scale, created
as the tachyon vanishes.  Since the
Bogoliubov coefficients are simply conjugated under time reversal, we 
expect a thermal state of temperature $1/\beta$, but this thermal bath
is created as the tachyon becomes strong.  We actually find a much stronger
source of backreaction.  We will begin by solving the dilaton fluctuation
equation (\ref{dileom3},\ref{dileom4}) 
in a truncated form; then we present numerical 
calculations of the strength of the backreaction.

\subsection{Solutions}

We are most interested in the time dependence of the dilaton equation of 
motion (\ref{dileom3},\ref{dileom4}), 
so we truncate the somewhat complicated spatial 
dependence (that is, the appearance of both spatial derivatives and spatial
positions in the differential equation).
Specifically, we will work with
\eq{dileom5}{\left[ \del_t^2-2V_t\del_t+k^2 +m^2 e^{\beta t}
\left(c_1+c_2\del_t\right)\right]\phi =0 \ ,}
where $t=X^0$ is the usual time coordinate.
We obtain this simplified equation of motion by setting $X^a$ to 
a small value in either (\ref{dileom3}) or (\ref{dileom4}), 
then treating $X^a$ as a constant while 
Fourier transforming to momentum space.  (In addition, we set $X^1=0$ and 
dropped $X^1$ derivatives in the lightlike gradient case.)  We
also ignore the subleading (nonexponential) time dependence in the tachyon
profile $T$.  We have also combined many of the constants into single variables
($m^2$, $c_1$, $c_2$) for notational convenience.  It is also helpful, 
of course, to rewrite the equation of motion in terms of $x=\beta t$, which
gives
\eq{dileomx}{\left[ \del_x^2+2V\del_x+k^2/\beta^2 +M^2 e^{\beta t}
\left(a+b\del_x\right)\right]\phi =0 \ .}
For reference, the constants appearing in (\ref{dileomx}) are
\bea
V&=& |V_t/\beta| = \alp V_t^2/2 = (D-26)/12\nonumber\\
M^2 & =& \gamma^2\mu^2(X^a X^a)/8\alp\beta^2\lb{parameters1}
\eea
In the lightlike case, 
\bea
a&=&4D-2\alp\beta^2 V_t -\frac{1}{4}\alp{}^2\beta^2 k^2-\left(
\frac{\alp{}^2}{(X^a)^2}\right)\left(\frac{1}{2}V_t^2+\frac{1}{2}N\beta V_t
\right)\nonumber  \\
b&=&\alp\beta^2 +\left(\frac{\alp{}^2}{(X^a)^2}\right)\left(
\frac{1}{4}N\beta^2-\frac{1}{4}(2N+1)\beta V_t\right)\ ,\lb{lightparams}
\eea
and, in the timelike case,
\bea
a&=&4D-2\alp\beta^2 V_t -\frac{1}{4}\alp{}^2\beta^2 k^2
-\frac{1}{4}\alp{}^2\beta^3 V_t+\frac{3}{16}\alp{}^2\beta^4+
\frac{1}{4}\alp{}^2\beta^2 V_t^2\nonumber\\
&&-\left(\frac{\alp{}^2}{(X^a)^2}\right)\left(\frac{3}{2}N\beta V_t 
+\frac{1}{2}N V_t^2+\frac{3}{4}N\beta^2\right)\nonumber  \\
b&=&\alp\beta^2+\frac{3}{8}\alp{}^2\beta^3 
+\left(\frac{\alp{}^2}{(X^a)^2}\right)\left(
\frac{1}{4}N\beta^2\right)\ .\lb{timeparams}
\eea

In our study, we will study a simplifying limit in which the parameters
$a$ and $b$ coincide for the lightlike and timelike tachyon cases.  First,
we work in the large $D$ limit (which is where the timelike tachyon background
is reliable anyway).  We also set $N\sim 1$ and $(X^a)^2\gg D\alp$.  In 
this limit, 
\eq{parameters2}{a=4D\ ,\ \ b=\alp\beta^2}
in either case; the most important
parameters for us will turn out to be $V=D/12$ and $a/b=D^2/6=24V^2$ in this
limit.  It is also important to note two features.
First is that $M^2$ may take either sign,
depending on the sign of the tachyon (recall that $\mu^2$ may be positive or
negative).  Second, note that the tachyon amplitude is controlled by 
$M^2 b$.  

The solution to (\ref{dileomx}) is 
\eq{dilsoln1}{\phi\left(x,\vec k\right) = e^{-Vx} 
\left[A e^{\nu x} M\left(\nu+
\frac{a}{b};1+2\nu;M^2 b e^x\right) +B e^{-\nu x} M\left(-\nu+
\frac{a}{b};1-2\nu;M^2 b e^x\right)\right]\ ,}
where $\nu = \sqrt{V^2-k^2/\beta^2}$.  The function $M$, 
also denoted ${}_1F_1$, is the 
Kummer or confluent hypergeometric function.  These solutions match to
positive and negative frequency modes in the far past, with frequency given
by $\nu$, though those modes are real exponentials at long wavelengths (real
$\nu$).  Both the solutions grow rapidly as $\exp[ M^2 be^x]$ in the future,
but the tachyon is no longer a perturbation of the background by then; we
can trust (\ref{dilsoln1}) only for $t< -(1/\beta)\ln M^2 b$.

\subsection{Source Term Estimates}
To study the backreaction of quantum fluctuations of the dilaton, we can
examine the dilaton equation of motion in the linear dilaton background.
Including the expectation of quantum fluctuations, this is 
\eq{backgroundeom}{V_\mu V^\mu+\Del^2\Phi -\del_\mu\Phi\del^\mu\Phi- 
\langle\kappa^2\del_\mu\phi\del^\mu\phi\rangle=0\ ,}
where $\Phi$ is the background dilaton plus fluctuation $\Phi=\Phi_0+
V_\mu X^\mu +\kappa\phi$.  ($\kappa$ is given by the $D$-dimensional
Planck scale since $\phi$ is of
canonical dimension while $\Phi$ is dimensionless.)
The quantum fluctuation yields a source term through the
last term in (\ref{backgroundeom}); backreaction will be important when
$\langle \kappa^2 (\del\phi)^2\rangle\gtrsim D/\alp\sim |V_\mu V^\mu|$

The expectation value $\langle \kappa^2 (\del\phi)^2\rangle$
is given by the average of the 
fluctuations on the length scale $1/V_t$, the only scale of the
cosmological linear dilaton background.  In particular, we choose the momentum
scale $V_t$ rather than $\beta$ because $V_t$, like the Hubble scale in 
cosmological backgrounds, is the scale at which fluctuations switch from
frozen or growing behavior to oscillatory behavior.  (In fact, short
wavelength fluctuations behave just like free massless fields, which do not
contribute to the source term at all.)  
The reader may wonder if we can trust our equation of motion (\ref{dileomx})
at wavenumbers beyond $1/\sqrt{\alp}$ when $|V_t|>1/\sqrt{\alp}$
since $\alp$ corrections should enter at that wavenumber. 
We will address this point below.

We see that we need to calculate
\eq{backreact1}{\langle \kappa^2(\del\phi)^2\rangle =\kappa^2
\int^{|V_t|}_0 \frac{dk}{(2\pi)^{D-1}}
k^{D-2}\left(k^2|\phi|^2-|\dot \phi|^2\right)\ .}
The momentum modes in this integral are precisely those with the real 
exponential behavior in the far past; in order to avoid large backreaction
at infinite wavelength in the far past, 
we require that the integration constant
$B=0$.  Then, for proper normalization, we should take $A=\sqrt{-i/\beta\nu}$.
(If we want to start in a vacuum state at the point where the string coupling
of the linear dilaton becomes small, we should have a small admixture of $B$,
but this should not change our results significantly.)

Defining 
\eq{phidef}{\phi(x) = 
\sqrt{-\frac{i}{\beta\nu}}e^{-Vx}\sigma(z)\ ,\ \ 
z=M^2 b e^x\ ,}
we find
\eq{backreact2}{\langle \kappa^2(\del\phi)^2\rangle =
\frac{\kappa^2\beta^D e^{-2Vx}}{(2\pi)^{D-1}}
\int^V_0 d\nu \left(V^2-\nu^2\right)^{(D-3)/2} \left(
\left(V^2-\nu^2\right)|\sigma|^2-\left|V\sigma-z\del_z\sigma\right|^2\right)
\ .}
To determine the extent of the backreaction when the tachyon becomes
important, we should study this integral at $z=\pm 1$.  This is easiest to
approximate for a large dimensionality $D$.  As discussed at the end of 
the last subsection, $V=D/12$ and $a/b=24V^2$ in that limit.  
Finally, the Planck scale is given by 
$\kappa^2\sim(2\pi)^{D-3}\alp{}^{(D-2)/2}e^{2\Phi_0}$, so we find
\eq{backreact3}{\langle \kappa^2 (\del\phi)^2\rangle\approx
\frac{g_s^2}{\alp} \left(\frac{24}{D}\right)^{D/2} I_D\ ,}
with $I_D$ the integral from (\ref{backreact2}).  Here,
$g_s$ is the string coupling at the time we study, when the tachyon 
amplitude $M^2 be^x$ reaches order unity.

We have integrated $I_D$ numerically for several values of the
dimensionality up to $D=480$ for both signs of the tachyon (\textit{i.e.},
$z=\pm 1$).
We find that $I_D$ grows much more quickly than $D^{D/2}$ in 
either case; in fact, even if we ignore the factors of $2\pi$ in the Planck
scale, the backreaction can still become large, since $I_D$ appears to
grow even faster than $(2\pi)^D D^{D/2}$.  For example, $I_D$ 
becomes larger than $(2\pi)^D D^{D/2}$ for $z=-1$ at $D\gtrsim 240$, while 
it is 167
orders of magnitude larger for $z=1$ at that dimensionality.  
These results are summarized in
table \ref{t:integral}.  We stress that the top row gives the best estimate
of the source term; the second row is given as a (very) conservative lower
limit.

\TABLE{\begin{small}
\begin{tabular}{|l|l|l|l|l|l|l|l|l|l|l|l|l|}
\hline
Dimension & & 48& 72 & 96 & 120 & 144 & 168 & 192 & 216 & 240 & 360 & 480\\
\hline
$\log(I_D(24/D)^{D/2})$ & $z=1$& 53& 87&123&160&198&237&277&317&358&571&792\\
& $z=-1$ & 19&40&54&74&95&118&140&163&187&315&452\\
\hline
$\log(I_D(24/4\pi^2 D)^{D/2})$ & $z=1$& 15&30&46&64&83&103&124&145&167&284&
409\\
& $z=-1$ & -20&-17&-22&-22&-20&-16&-13&-9&-4&28&69\\
\hline
\end{tabular}\end{small}
\caption{\lb{t:integral} The (base-10 log of the) 
integral $I_D$ calculated with $V=D/12$ 
and $a/b=D^2/6$ for a range of dimensionalities.}
}

As mentioned above, a cautious reader will not trust equation (\ref{dileomx})
for wavenumbers $k\gtrsim 1/\sqrt{\alp}$ because string worldsheet corrections
should become important at that scale.  Therefore, to be cautious, we 
consider summing over only momenta up to $k=1/\sqrt{\alp}$.  There are two 
points to make before presenting the results.  First, $V_t$ is smaller than
the string scale as long as $D\leq 182$, so the correct results are those
of (\ref{t:integral}) for those dimensions.  Second, it may not be necessary
to cut off the integral at $k\sim 1/\sqrt{\alp}$ if we are only concerned with
time dependence, since the time derivative of the dilaton fluctuation
is always smaller than the linear dilaton gradient, the natural scale of the
problem.  In any case, though, to be cautious, we present results at some
large dimensions with a upper momentum cut-off of $k= 1/\sqrt{\alp}$.
This just
changes the upper integration limit of (\ref{backreact1}) to 
$1/\sqrt{\alp}$.  In equation (\ref{backreact2}), the lower integration limit
becomes $\nu=\sqrt{V^2-1/\alp\beta^2}=\sqrt{V^2-V/2}\approx V-1/4$.
Adding this lower limit restricts the backreaction a great deal.  In the
case of negative tachyon sign ($z=-1$), we find that the backreaction is 
neglible for large $D$ (and decreasing as $D$ increases).  However, the 
negative sign tachyon still gives a large (and increasing) source term at
large $D$.  These results are summarized in table \ref{t:integral2}.

\TABLE{\begin{small}
\begin{tabular}{|l|l|l|l|l|l|l|l|l|l|}
\hline
Dimension & &  192 & 216 & 240 & 360 & 480 & 540 &600&660\\
\hline
$\log(I_D (24/D)^{D/2})$ & $z=1$& 80&90&100&151&201&226&251&277\\
& $z=-1$ & -55&-60&-67&-100&-133&-149&-166&-183\\
\hline
\end{tabular}\end{small}
\caption{\lb{t:integral2} The integral $I_D$ calculated with $V=D/12$ 
and $a/b=D^2/6$ for a range of dimensionalities.  In this case, the lower
limit of the integral is $V-1/4$.}
}
\comment{
Finally, since the lightlike tachyon background is exact even for 
dimensionality near critical, we have evaluated $I_D$ for a $D=27,28$ 
without using the large dimension approximation.  We find that the
backreaction is neglible in these cases.  }

\subsection{Discussion}
\FIGURE{
\includegraphics[scale=0.6]{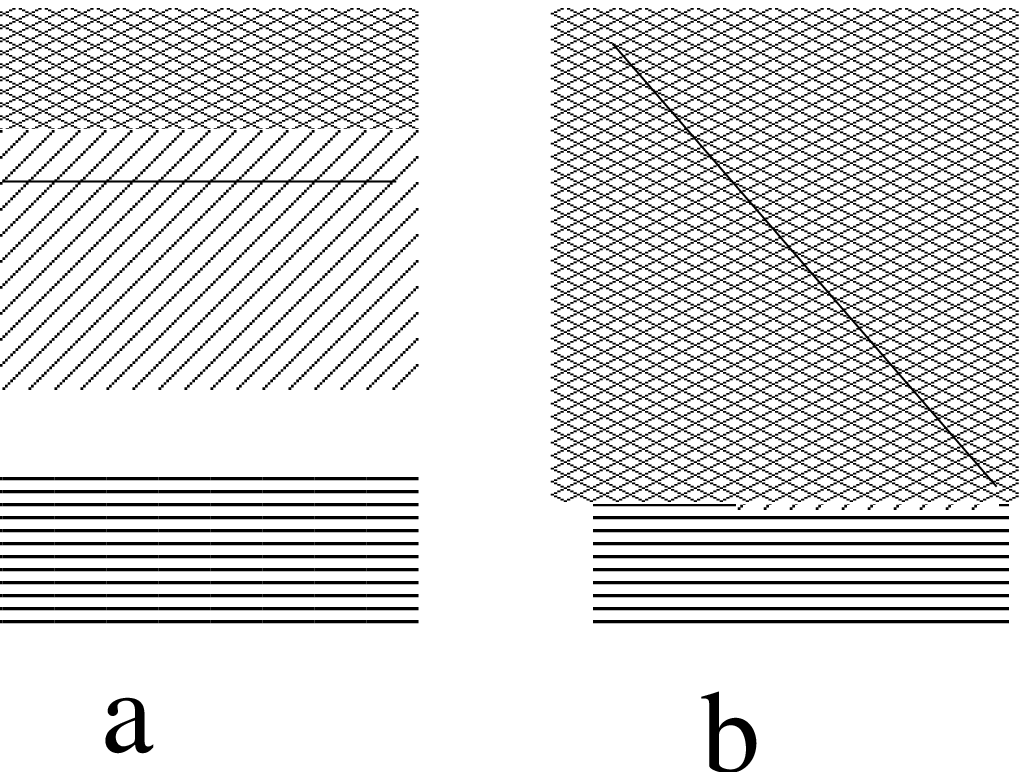}
\caption{\lb{f:back}Regions of strong coupling (horizontal shading), strong
tachyon (cross-hatched shading), and possible strong backreaction (diagonal
shading) for timelike (a) and lightlike (b) tachyon gradients.}}

From the above calculations,
it appears that the only way to avoid (severe) backreaction effects at
large $D$ is
to tune the string coupling to be extremely small at the time the tachyon
becomes strong.  We stress that backreaction can be quite significant well
into the region of perturbative string coupling.  The backreaction is, 
in fact, perturbative in the string coupling, but the coefficient of $g_s^2$
is much larger than unity.

In the case of timelike tachyon gradient, this is not 
burdensome, since we usually assume the strong coupling regime
to be far in the past in order to use weakly coupled string theory.  
Keeping the backreaction under control just means that we push the strongly
coupled region even further into the far past.  As a caution, though,
just due to the large numbers involved,
it seems like backreaction may be important even at times well before 
the tachyon background becomes strong. In this case, we can ask
whether any effects can propagate from a region of significant backreaction.
Since any string matter generated through backreaction only generates
effects of order $g_s$ or higher, we expect that having the strong 
tachyon region in sufficiently weak string coupling will render backreaction
effects negligible.  However, it is possible to imagine a scenario in which
the backreaction (in a region of both weak coupling and weak tachyon)
could either cause the dilaton to grow again or accelerate the growth of
the tachyon, leading to difficulties with the usual picture.  Of course, 
it seems that ensuring the string coupling is sufficiently weak for a very
long time before the tachyon becomes important can control backreaction
effects.  Therefore, it is important to
stress that the strong coupling era must end quite some time before the
tachyon becomes strong.
As a reference, in table \ref{t:times}, we list the number of string times
the strong coupling and strong tachyon regions must be separated by (at
a bare minimum) to avoid strong backreaction effects in some cases.

The situation is slightly more complicated in the case of a lightlike
tachyon gradient.  In that case, there is always
a region where the tachyon is large and the dilaton is not extremely
small.  In other regions where the tachyon is strong, of course, the dilaton 
\textit{will} be extremely small, so backreaction will be negligible; the
main question is whether any effects can propagate from the region of
important backreaction, as discussed above.  Once again, we expect that
sufficiently small string coupling will control the backreaction.
The regions of significant backreaction are
summarized in figure \ref{f:back}.

\TABLE{\begin{small}
\begin{tabular}{|l|l|l|l|l|l|l|l|l|l|l|l|l|}
\hline
Dimension & &48& 72 & 96 & 120 & 144 & 168 & 192 & 216 & 240 & 360 & 480\\
\hline
$\Delta t$ & $z=1$ &15&17&18&18&19&19&20&20&21&22&23\\
& $z=-1$& 5&8&8&8&9&10&10&10&11&12&13\\
\hline
\end{tabular}\end{small}
\caption{\lb{t:times} The number of string times needed to separate the strong
coupling and strong tachyon regions in order to avoid backreaction at the
time of strong tachyon.  These results use the calculations in table
\ref{t:integral}.}}

Of course, we have so far only carried out a rough calculation of the 
backreaction in the bosonic string.  
However, the strength of our results indicates that backreaction
is a serious issue that may complicate the picture of closed string tachyon
condensation, and these concerns will certainly also arise in the heterotic
case. Sadly, backreaction may spoil the clean picture of the
closed string tachyon as smoothly annihilating dimensions of spacetime, at
least in some regions.  To gain a more complete understanding of closed
string tachyon condensation, we will need to understand backreaction due
to the production of massless string modes in the tachyon background.
In the meanwhile, our results serve as a caveat for interpreting 
tachyon condensation.

\acknowledgments
I would like to thank Robert Brandenberger and Sugumi Kanno for collaboration
on \cite{arXiv:0706.1104}, which led to this project. I would also like to
thank Robert Brandenberger and Guy Moore for 
very useful discussions as well as Albion Lawrence for comments.  I am 
especially grateful to Simeon Hellerman for explaining some subtleties of the
calculation in section \ref{s:conformal}.

This work is supported by NSERC through the Discovery Grant
program and in part by the Institute for Particle Physics and the Perimeter
Institute.

\appendix
\section{BRST Quantization in Linear Dilaton Background}\label{a:lineardilaton}
For reference, we provide here the BRST quantization of a closed string
at the tachyonic and massless levels in a linear dilaton background.  
The results are necessary for us to simplify the string amplitude calculated
in section \ref{s:amplitude}.  We largely follow the discussion for the 
Minkowski string in \cite{Polchinski:1998rq}.

On the complex plane, the holomorphic and antiholomorphic worldsheet stress
tensors are just given by the (Euclideanization of) \ref{stressflat} in
the linear dilaton background.  Therefore, the first few Virasoro generators
in the scalar sector are
\bea
L^X_0 &=&\frac{1}{2\pi\alp}\oint dz\, zT(z) = \frac{\alp}{4}p^2
+\sum_{n=1}^\infty
\alpha_{-n}\alpha_n+i\frac{\alp}{2}V_\mu p^\mu \nonumber \\
L_1^X &=& \frac{1}{2\pi\alp}\oint dz\, z^2T(z) =
\frac{1}{2}\sum_{n=-\infty}^\infty\alpha_{1-n}\alpha_n+i\sqrt{2\alp}
V_\mu \alpha_1^\mu\nonumber\\
L_{-1}^X&=&\frac{1}{2\pi\alp}\oint dz\, T(z) = 
\frac{1}{2}\sum_{-\infty}^\infty\alpha_{-1-n}\alpha_n\ .\label{virasoro}
\eea
The antiholomorphic generators are just the complex conjugates, as usual.  We
also need the contribution to the Virasoro operators from the ghosts, which 
we can copy from \cite{Polchinski:1998rq}.  These are
\eq{ghostvir}{L_0^g=\sum_{n=-\infty}^\infty n\! \normal{b_{-n} c_n}\, ,\ 
L_1^g=\sum_{n=-\infty}^\infty (2-n)\!\normal{b_n c_{1-n}} \, ,\ 
L_{-1}^g=-\!\!\!\sum_{n=-\infty}^\infty(2+n)\!\normal{b_n c_{-1-n}} .}
Here the normal ordering symbol stands for creation-annihilation normal 
ordering, unlike in the main text.

The BRST operator for the linear dilaton is then
\eq{BRST}{Q_B=\sum_n \left(c_{n}L^X_{-n}+\t c_n\t L_{-n}^X\right) 
+\sum_{m,n}\frac{m-n}{2}\left(\normal{c_mc_n b_{-m-n}} +\normal{\t c_m\t c_n 
\t b_{-m-n}}\right) -c_0 -\t c_0\ .}
A physical state $\ket{\psi}$ of the string must be in the cohomology of $Q_B$
and satisfy $b_0\ket{\psi}=\t b_0\ket{\psi}=0$ and $L_0\ket\psi=
\t L_0\ket\psi=0$.  We can ensure the $b_0$ condition just by taking the 
appropriate ghost ground state; since $L_0 =\{Q_B,b_0\}$, the $L_0$ conditions
follow.  In practice, however, we'll find it instructive to examine the
complete $L_0$ condition, which works out to be
\eq{L0condition}{L_0\ket\psi=0\Rightarrow\ \ p^2+2iV\cdot p=
-\frac{4}{\alp}(N-1)\ ,}
where $N$ is the total holomorphic matter plus ghost oscillator excitation
number.  The antiholomorphic sector gives the same condition with $N\to \t N$,
but level matching requires $N=\t N$.

At the tachyonic level, the state of the string can only be $\ket{0;k}$
with $k^2+2iV\cdot k=4/\alp$.  In addition, 
\eq{tachyonicQB}{Q_B\ket{0;k} = \left(c_0L_0+\t c_0 \t L_0\right)\ket{0;k}=0}
just because of the mass-shell condition.

At the massless level ($k^2+2iV\cdot k=0$), the most general state is
\bea
\ket\psi & =& \left(e_{\mu\nu}\alpha_{-1}^\mu\t\alpha_{-1}^\nu +f_\mu
\alpha_{-1}^\mu \t b_{-1}+\t f_\mu b_{-1}\t\alpha_{-1}^\mu 
+g_\mu\alpha_{-1}^\mu\t c_{-1} +\t g_\mu c_{-1} \t \alpha_{-1}^\mu\right.
\nonumber\\
&&\left. +hb_{-1}\t c_{-1}+\t h c_{-1}\t b_{-1}+\beta b_{-1}\t b_{-1}
+\gamma c_{-1}\t c_{-1}\right)\ket{0;k}\ .\label{massless}\eea
The BRST operator on this state is
\bea
Q_B\ket\psi &=& \sqrt{\frac{\alp}{2}}\left[e_{\mu\nu}(k+2iV)^\mu c_{-1}
\t\alpha_{-1}^\nu +e_{\mu\nu}(k+2iV)^\nu \alpha_{-1}^\mu \t c_{-1}
+f\cdot(k+2iV) c_{-1}\t b_{-1}+f\cdot\alpha_{-1} k\cdot\t\alpha_{-1}\right.
\nonumber\\
&&\left.+\t f\cdot (k+2iV) b_{-1}\t c_{-1}+\t f\cdot\t\alpha_{-1}^\mu 
k\cdot\alpha_{-1}+g\cdot(k+2iV)c_{-1}\t c_{-1}+\t g\cdot(k+2iV)c_{-1}\t c_{-1}
\right.\nonumber\\
&&\left. +hk\cdot\alpha_{-1}\t c_{-1} +\t h c_{-1}k\cdot\t\alpha_{-1}
+\beta k\cdot \alpha_{-1} \t b_{-1}+\beta b_{-1}k\cdot\t\alpha_{-1}\right]
\ket{0;k}\ .\label{masslessQB}
\eea
If $\ket\psi$ is BRST-closed, then we must have
\bea e_{\mu\nu}(k+2iV)^\mu+\t hk_\nu&=&0\ ,\ \ e_{\mu\nu}(k+2iV)^\nu
+hk_\mu =0\ ,\ \beta=0\ ,\nonumber\\ 
f_\mu=\t f_\mu&=&0\ ,\ g\cdot (k+2iV)=
\t g\cdot(k+2iV)\ .\label{massless-closed}\eea

The general BRST-exact state at this level is of the form (\ref{masslessQB})
with primed coefficients.  By choosing $f'_\mu$ and $\t f'_\mu$, we can 
therefore gauge away $h$ and $\t h$ (compare terms in (\ref{massless}) and
(\ref{masslessQB})).  By choosing $e'_{\mu\nu}$, we can gauge away $g_\mu$
and $\t g_\mu$.  Finally, by choosing $g'_\mu$ and $\t g'_\mu$, we can 
gauge away $\gamma$.  We are required to choose $\beta'=0$ to maintain the
condition $f_\mu=\t f_\mu=0$.  Once we have made these choices, we can make
a further transformation with $f''_\mu$ and $\t f''_\mu$ as long as
$f''\cdot(k+2iV)=\t f''\cdot(k+2iV)=0$, which shifts $e_\mu\nu$.  We are left
with the following state, conditions, and gauge equivalence:
\bea
\ket\psi &=& e_{\mu\nu}\alpha_{-1}^\mu\t\alpha_{-1}^\nu \ket{0;k}\nonumber\\
0&=&e_{\mu\nu}(k+2iV)^\mu = e_{\mu\nu}(k+2iV)^\nu\nonumber\\
e_{\mu\nu}&\simeq& e_{\mu\nu}+\sqrt{\frac{\alp}{2}}f''_\mu k_\nu
+\sqrt{\frac{\alp}{2}}k_\mu \t f''_\nu\ .\label{masslesscohomology}\eea
The polarization tensor $e_{\mu\nu}$ can then be separated into a symmetric
graviton, 2-form potential, and dilaton trace parts, as normal.

\bibliographystyle{h-physrev4}\bibliography{weak}
\end{document}